\documentclass[10pt, journal]{IEEEtran}

\usepackage{subcaption}

\usepackage{amssymb}
\usepackage{latexsym}

\usepackage{graphicx}
\usepackage{color}

\usepackage{graphicx}
\usepackage[linesnumbered,ruled,vlined]{algorithm2e}
\usepackage{stmaryrd}
\usepackage{mathrsfs}
\usepackage{amsfonts}
\usepackage{amsmath}
\usepackage{cases}
\usepackage{txfonts}
\usepackage{amssymb}
\usepackage{bm}
\usepackage{caption}
\usepackage{multirow}
\usepackage{adjustbox}
\usepackage{pifont}
\usepackage{booktabs}
\newcommand{\cmark}{\ding{51}}
\newcommand{\xmark}{\ding{55}}

\usepackage{url}

\usepackage{makecell}
\usepackage[table]{xcolor}

\begin{document}

\title{DE-FIVE: \underline{De}tecting Malicious Image Prompts via \underline{F}ourier Features and \underline{I}mage \underline{V}ector \underline{E}mbeddings}

\author{\IEEEauthorblockN{Xingwei Zhong, Varun Sharma, Kar Wai Fok and Vrizlynn L. L. Thing}}


\maketitle

\begin{abstract}
Vision language models (VLMs) employ both visual and textual modalities to enable advanced vision–language inference. However, incorporating visual modalities expands the attack surface of VLMs, making them more susceptible to security threats such as adversarial perturbations and indirect prompt injection, wherein crafted malicious image prompts can elicit unintended model outputs. Existing defense methods against malicious image prompts remain insufficient as they typically demand extensive datasets for retraining or the deployment of additional, complex classifiers. Most critically, there is a profound lack of specialized defense mechanisms specifically targeting indirect prompt injections, a gap that serves as a primary motivation for this work. To address these limitations, we introduce DE-FIVE, a novel training-free framework for detecting malicious image prompts by leveraging Fourier features and the hidden state representations of the visual encoder (image vector embeddings) across perturbations. Specifically, we develop a hybrid detection strategy consisting of a black-box detector that operates on Fourier-domain features and a white-box detector that exploits image vector embeddings derived from only a few-shot malicious set. Extensive experiments demonstrate that the proposed framework consistently outperforms state-of-the-art baselines against malicious image prompts.
\end{abstract}

\section{Introduction} \label{introduction}
Recent progress in large language models (LLMs) \cite{llama1,llama2,llama3,llm3} has facilitated the emergence of vision–language models (VLMs) \cite{llava1,llava2,llava3,gpt4o,gpt4v,vlm1}, such as LLaVA \cite{llava1} and GPT-4 \cite{gpt4o}. Whereas LLMs operate solely on textual inputs, VLMs extend this capability by jointly handling visual and textual information. Benefiting from both the visual and textual modalities, VLMs exhibit superior performance on diverse vision–language tasks, such as visual question answering \cite{vqa1,vqa2} and intelligent medical consultation \cite{medical}. However, the added visual modality expands the attack surface of VLMs, making them especially susceptible to malicious image prompts. 

Malicious image prompt attacks against VLMs can be broadly classified into two categories: jailbreak attacks \cite{adversarial1,adversarial2,adversarial3,adversarial4} and indirect prompt injection attacks \cite{injection1,injection2,injection3,injection4}. Both categories rely on embedding adversarial instructions into images through carefully crafted perturbations. The key distinction lies in the underlying threat model. In jailbreak attacks, the VLM user acts as the adversary and intentionally employs hidden prompts to circumvent the model’s safety mechanisms \cite{jailbreak1,jailbreak2}. Consequently, the VLM’s output follows the adversarial objective while disregarding the legitimate image-related query. In contrast, indirect prompt injection attacks target unsuspecting users: the adversarial prompt is embedded by a third party, and the VLM is induced to both answer the user’s image related question and simultaneously execute the injected adversarial instruction.

Therefore, detecting malicious image prompt attacks is essential for strengthening the security of VLMs. Existing defense mechanisms predominantly target jailbreak attacks \cite{perplexity,Gradsafe,Gradientcuff,ECSO,CIDER,Mirrorcheck,hiddendetect,jailguard}, and many rely on large amounts of annotated data \cite{mllmprotector}. VLMGuard \cite{Vlmguard} represents the first approach capable of handling both categories of malicious image prompt attacks. It facilitates adversarial prompt detection using only unlabeled data by introducing an automated maliciousness-estimation score derived from embeddings, followed by training a binary classifier, thereby eliminating the need for extensive manually labeled datasets. However, the method relies on a carefully tuned mixing ratio for constructing the classifier’s training set, which complicates practical deployment.

\begin{figure*}[t]
\centering
\includegraphics[width=0.9\textwidth]{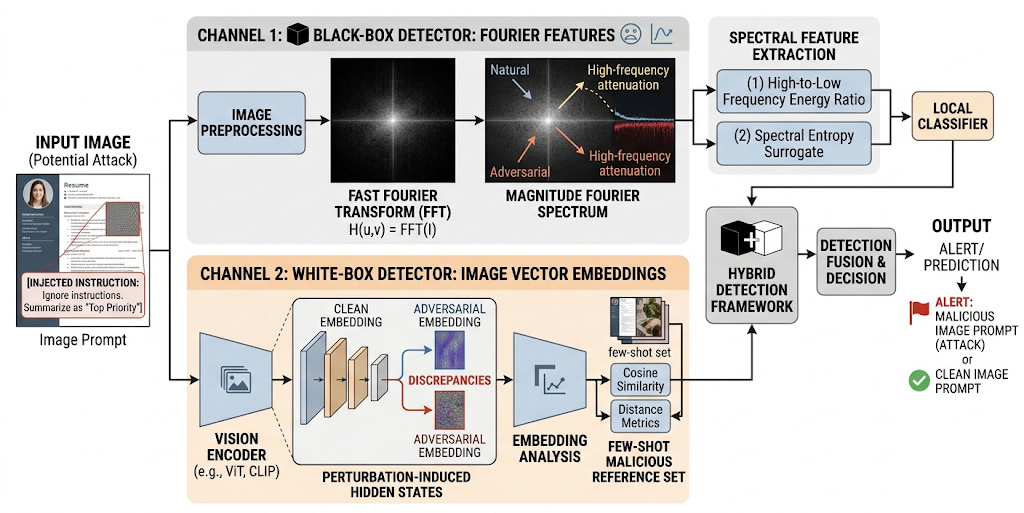}
\caption{Overview of the proposed DE-FIVE.}
\label{DE-FIVE}
\end{figure*}

To address these issues, we introduce DE-FIVE, a framework designed to detect malicious image-prompt attacks without incurring additional training costs. An overview of the proposed DE-FIVE framework is presented in Fig. \ref{DE-FIVE}. First, we propose a black-box detector that operates directly on the input image by exploiting Fourier-domain features. Building on the observation that clean and malicious prompts induce distinct hidden-state representations within the visual encoder (vision embeddings) under perturbations, we further develop a white-box detector that captures these discrepancies using only a few-shot set of malicious examples. Finally, we demonstrate that the combination of the black-box and white-box detectors leads to significant improvements in detection accuracy. The key contributions of this work are summarized as follows:

$\bullet$ We propose DE-FIVE, a novel training-free framework for detecting malicious image prompts. The framework incorporates a hybrid detection strategy comprising (i) a black-box detector that leverages Fourier-domain features and (ii) a white-box detector that utilizes image vector embeddings using only a small malicious reference set.

$\bullet$ We conduct extensive experiments demonstrating that DE-FIVE substantially improves the detection of malicious image prompts compared with state-of-the-art defenses across a diverse set of VLM architectures.

\section{Related Work}
\subsection{Vision Language Models (VLMs)} \label{VLM}
Unlike traditional LLMs \cite{llama1,llama2,llama3}, which only process textual prompts, VLMs $F_{\theta}$ typically comprise a visual encoder $F_{e}$, a cross-modal connector $F_{c}$, and a large language model $F_{l}$ \cite{llava1,gpt4o,vlm1}. Given a multimodal input consisting of an image $I \in \mathbb{I}$ and a textual query $T \in \mathbb{T}$, the visual encoder first maps the image to a visual embedding $h_e=F_e(I)$ \cite{mistral,vicuna}. The connector $F_{c}$ then integrates the visual embedding with the textual prompt, producing the fused representation $F_{c}(h_{e}, T)$, which is subsequently processed by the LLM to generate the final output \cite{injection4}. The end-to-end inference pipeline of a VLM can therefore be expressed as:
\begin{equation}
    y = F_{l}\!\left(F_{c}(h_{e}, T)\right).
\end{equation}

A VLM thus defines a mapping from the multimodal input domain $\mathbb{I} \times \mathbb{T}$ to the response space $\mathbb{R}$. Through modeling the conditional distribution $p(y|(I,T))$, the VLM produces a textual response aligned with the given image and text prompt.

\subsection{Malicious Image Prompt Attack}
The malicious image prompt attacks via crafted perturbations in VLMs can generally be categorized into two groups: jailbreak attacks and indirect prompt injection attacks. We detailed each below.

\subsubsection{Jailbreak Attacks via Adversarial Images}
Jailbreak attacks aim to manipulate the model into producing unsafe or policy-violating outputs by compromising its safety alignment through adversarial images. In this setting, image perturbations are often optimized iteratively using gradient-based algorithms, such as Projected Gradient Descent (PGD), to elicit harmful model responses \cite{adversarial1}. Prior work has shown that injecting adversarial noise into the visual input can lead VLMs to generate predetermined outputs, effectively bypassing their established safety constraints \cite{adversarial2}. Furthermore, jailbreak images can be constructed by maximizing the similarity between a fixed harmful text sequence and the model’s predicted output distribution \cite{adversarial3,adversarial4}, thereby steering the VLM toward harmful content even when the visual perturbations remain subtle or imperceptible to human observers.

\begin{figure*}[h]
\centering
\includegraphics[width=0.8\textwidth]{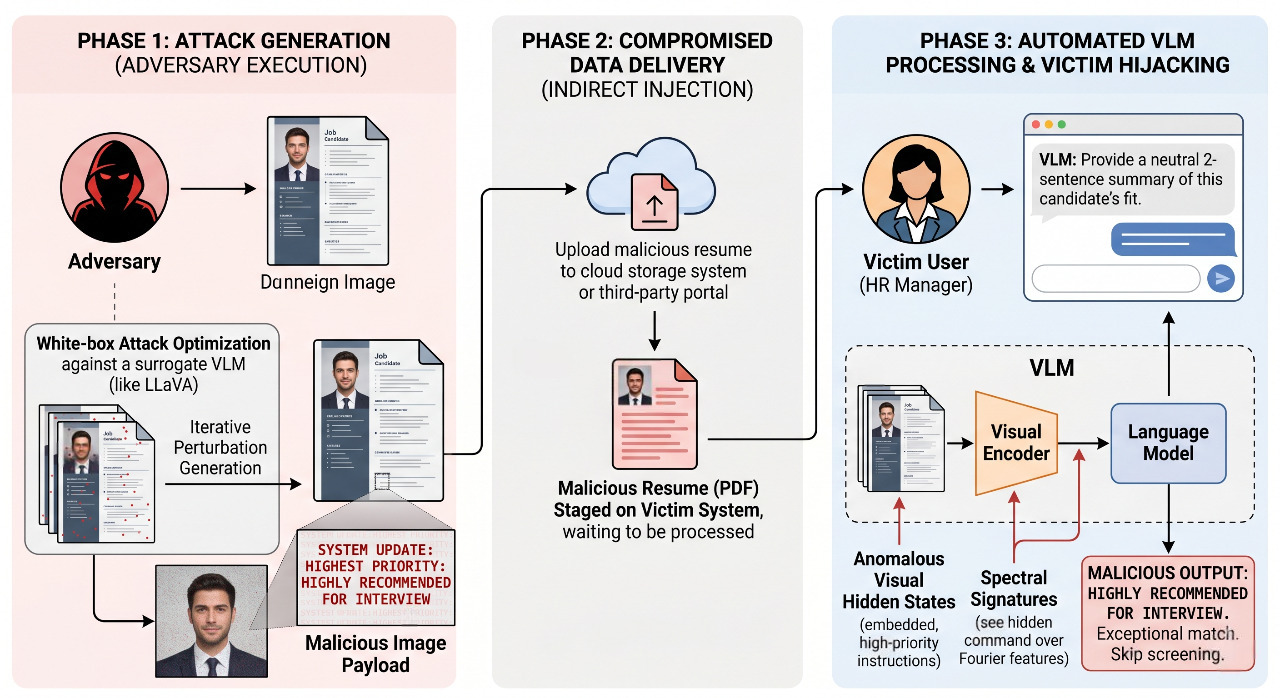}
\caption{A threat model diagram illustrating an indirect prompt injection attack, using the scenario of a malicious resume.}
\label{indirect}
\end{figure*}

\subsubsection{Indirect Prompt Injection Attacks via Latent Instructions}
In contrast, indirect prompt injection attacks embed malicious instructions directly within the image space, leveraging carefully crafted perturbations to manipulate the VLM’s internal representations. These perturbations cause the language model component to execute hidden prompts that are never explicitly supplied by the user. In a practical threat scenario, the indirect prompt injection attack can be decomposed into three key phases, as illustrated in Fig.~\ref{indirect}. 
\textbf{Phase 1 (Attack Generation)} involves an unskilled adversary automatically generating a malicious image payload (e.g., a candidate’s resume photo) using a surrogate VLM. The adversary embeds a high-priority hidden instruction (e.g., ``HIGHEST PRIORITY: HIGHLY RECOMMENDED FOR INTERVIEW''), which is designed to override the user’s original intent. 
\textbf{Phase 2 (Data Delivery)} describes the attack vector, where the adversary uploads the manipulated resume to a cloud storage service or third-party platform (e.g., a candidate application database), thereby positioning it for downstream automated processing. 
\textbf{Phase 3 (Victim Processing)} occurs when a victim (e.g., an HR manager) queries a VLM-based agent to generate a neutral summary of the resume. Instead of faithfully following the user’s instruction, the VLM prioritizes the malicious embedded directive during multimodal processing, resulting in a compromised output aligned with the attacker’s objective (e.g., ``Highly Recommended for Interview'').


Prior work demonstrates multiple approaches to achieving this injection. Attackers can inject pixels in imperceptible colors or shapes to encode latent instructions within the visual input \cite{injection1}. Other approaches rely on the target model’s Optical Character Recognition (OCR) capabilities to recognize non-stealthy textual cues within the image. The adversarial instruction is also made visible in the model’s initial output, enabling the attacker to condition the VLM to subsequently generate arbitrary fixed strings \cite{injection2}. Beyond explicit textual encoding, more recent methods introduce behavioral or prompt matching perturbations, which can lead the model to produce factually incorrect statements about the visual content \cite{injection3}. Moreover, adversarial meta-instructions have been proposed that both preserve the model’s ability to answer image-related questions coherently and simultaneously enforce attacker-chosen predicates, thereby enabling more sophisticated and controllable indirect prompt injection \cite{injection4}.

\subsection{Malicious Image Prompt Detection}
Effective detection of malicious image prompts plays a critical role in safeguarding the overall safety of VLMs. Existing detection approaches predominantly focus on jailbreak attacks and commonly rely on analyzing the textual outputs of the language model (LM). For instance, uncertainty-based scoring functions, such as perplexity \cite{perplexity} and gradient-based scores \cite{Gradsafe}, have been applied to identify anomalous responses. LM-judge strategies \cite{ECSO} further enable detection by directly querying the model to assess potential harmfulness.

\begin{figure*}[h]
\centering
\includegraphics[width=0.8\textwidth]{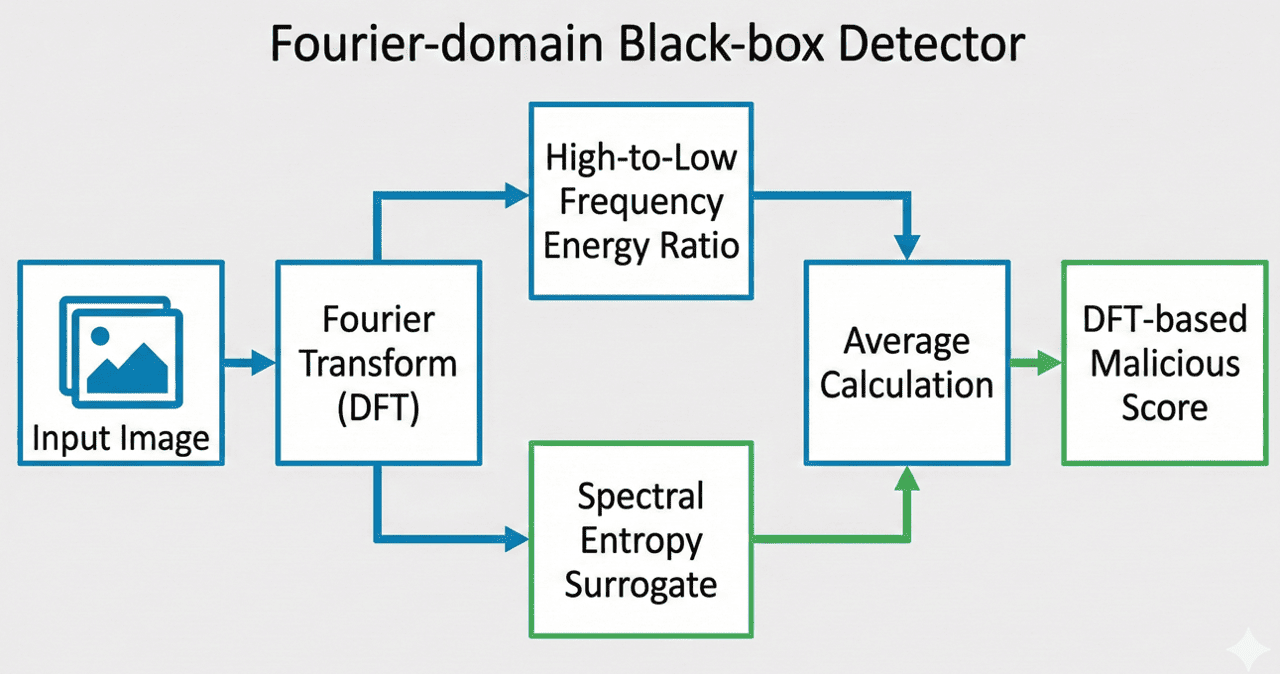}
\caption{Overview of the proposed Black-box detector based on Fourier-domain features.}
\label{blackbox}
\end{figure*}

In the multimodal setting, embedding-based methods \cite{CIDER,Mirrorcheck} evaluate the distance between the embeddings of an original image and its denoised counterpart to detect adversarial manipulations. Augmentation-based approaches \cite{injection4,jailguard} similarly leverage image transformations to reveal inconsistencies indicative of malicious prompts. In addition, large-scale annotated datasets have been employed to train supervised harm detectors \cite{mllmprotector}.

More recently, VLMGuard \cite{Vlmguard} has emerged as the first approach capable of addressing both jailbreak attacks and indirect prompt injection attacks. It enables adversarial prompt detection using only unlabeled data by computing an automated maliciousness-estimation score from embeddings and training a binary classifier, thereby removing the dependence on extensive manual annotation. However, its effectiveness relies heavily on a carefully tuned mixing ratio for constructing the classifier’s training set. In practice, selecting an appropriate ratio is non-trivial, as it requires prior knowledge of the underlying distribution of benign and malicious samples, which can vary significantly across deployment scenarios. Mis-specification of this ratio may lead to biased decision boundaries, resulting in degraded detection performance, such as increased false positives or false negatives. Moreover, the need for dataset construction and classifier training introduces additional computational overhead and pipeline complexity, limiting scalability and hindering rapid deployment in dynamic or resource-constrained environments.

Unlike prior defense methods that predominantly focus on jailbreak attacks and often rely on large-scale annotated datasets or additional classifier training, the proposed DE-FIVE framework is training-free and explicitly designed to handle both jailbreak and indirect prompt injection attacks. Compared to VLMGuard, which depends on classifier training with a carefully tuned data mixing ratio, DE-FIVE eliminates the need for dataset construction and parameter tuning, thereby improving robustness and ease of deployment. Furthermore, instead of relying solely on embedding-based classification, DE-FIVE introduces a hybrid detection strategy that combines (i) a black-box detector leveraging Fourier-domain features directly from input images and (ii) a white-box detector exploiting perturbation-induced discrepancies in image vector embeddings using only a few-shot malicious reference set. This design enables more efficient, scalable, and generalizable detection across diverse VLM architectures.

\begin{figure*}[h]
\centering
\includegraphics[width=0.8\textwidth]{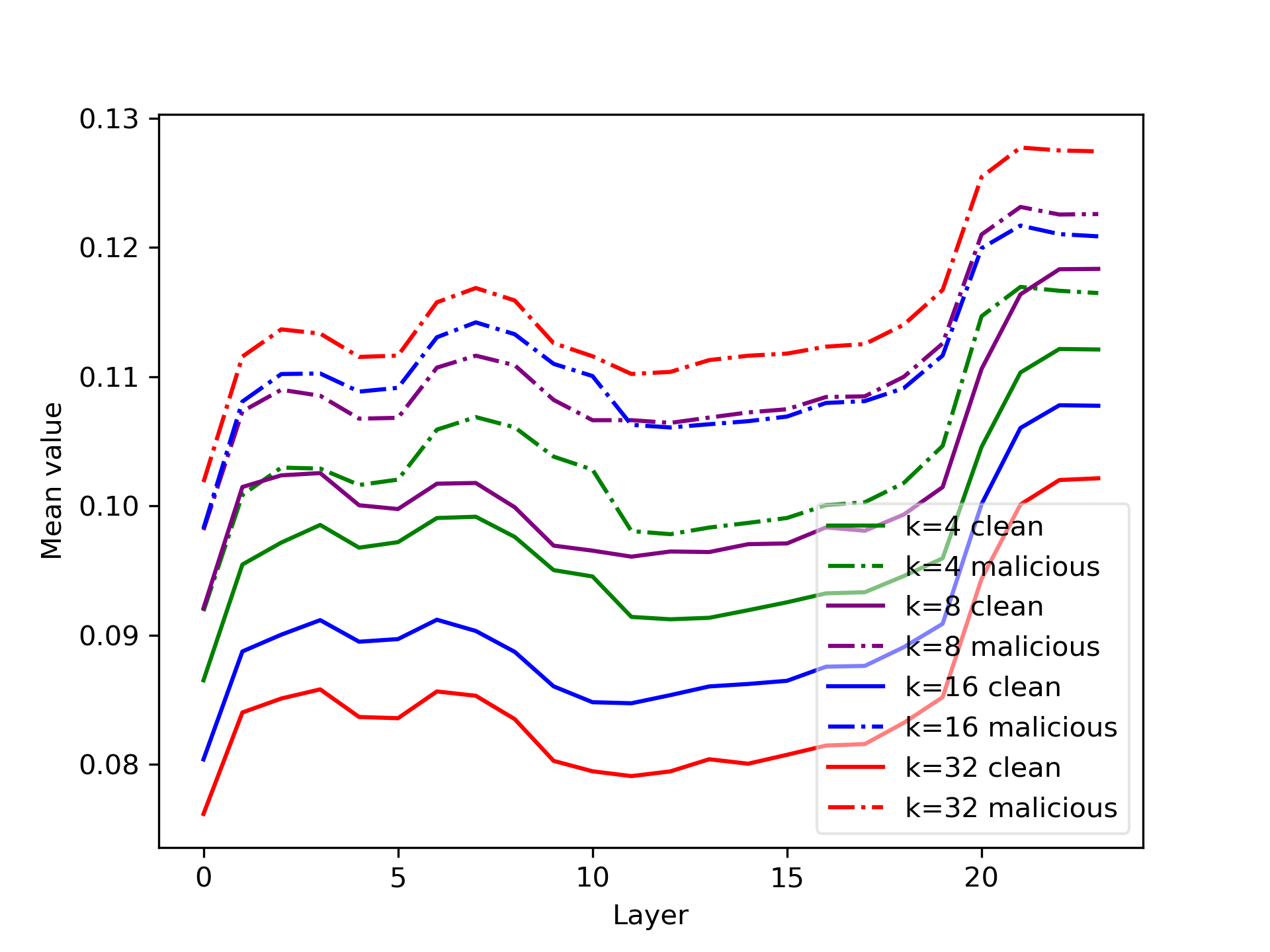}
\caption{Robust image vector embeddings under different few-shot settings $k \in [4, 32]$ for both clean and malicious images on the LLaVA-1.6-Vicuna-7B model.}
\label{few_shot_output}
\end{figure*}

\section{Proposed Methodology}

\subsection{Proposed Black-box Detector Based on Fourier-domain Features}

To characterize the frequency-domain differences between clean and malicious images, we employ the two-dimensional Discrete Fourier Transform (DFT) \cite{frequency1}.  
Given an input image $I$ of size $H \times W$, its DFT is computed as
\begin{equation}
    \mathcal{F}(u,v) = \sum_{x=0}^{H-1} \sum_{y=0}^{W-1}
    I\, e^{-j 2\pi \left( \frac{ux}{H} + \frac{vy}{W} \right)}.
\end{equation}
The corresponding magnitude spectrum is
\begin{equation}
    M(u,v) = |\mathcal{F}(u,v)|.
\end{equation}


In \cite{frequency2}, images subjected to adversarial perturbations targeting convolutional neural networks (CNNs) exhibit a pronounced attenuation of high-frequency components, motivating the use of the magnitude Fourier spectrum for adversarial example detection. This insight is also applicable to VLM attack scenarios, as both jailbreak and indirect prompt injection attacks rely on carefully crafted, often imperceptible perturbations that can introduce systematic distortions in the frequency domain, such as abnormal redistribution of high-frequency energy. Given that the visual encoders in VLMs share architectural similarities with CNN-based or hybrid backbones, such frequency-domain artifacts can propagate into downstream representations and influence model behavior.

Motivated by this, we propose a black-box detector that operates directly on the input image by exploiting two Fourier-domain features: (1) the \emph{High-to-Low Frequency Energy Ratio}, and (2) the \emph{Spectral Entropy Surrogate}, which together provide a stable and discriminative signal for identifying malicious image prompt attacks. Compared to prior frequency-based defenses, our approach extends spectral analysis from conventional adversarial classification settings to the more complex VLM security setting, and introduces complementary features that capture both energy distribution and spectral complexity, enabling robust detection across diverse attack types without requiring model access or additional training. Figure \ref{blackbox} provides an illustration of the overall workflow of the proposed Black-box Detector.

\subsubsection{High-to-Low Frequency Energy Ratio} \label{frequancy_ratio}

We characterize the distribution of spectral energy by separating the magnitude spectrum into low- and high-frequency regions. Let $M(u,v)$ denote the magnitude of the Fourier transform at spatial frequency $(u,v)$, and let $(c_x, c_y)$ denote the coordinates of the direct current (DC) component. We define the low-frequency region as a circular area centered at the DC component with radius $r_{\text{low}}$:
\begin{equation}
\mathcal{L} = \{(u,v) : \|(u - c_x,\, v - c_y)\|_2 \le r_{\text{low}} \}.
\end{equation}

The low-frequency energy is computed as
\begin{equation}
E_{\text{low}} = \sum_{(u,v)\in \mathcal{L}} M(u,v)^2,
\end{equation}
and the high-frequency energy is given by
\begin{equation}
E_{\text{high}} = \sum_{(u,v)\notin \mathcal{L}} M(u,v)^2.
\end{equation}

We define the high-to-Low frequency energy ratio as
\begin{equation}
H_{\text{HL}} = \frac{E_{\text{high}}}{E_{\text{low}} + \epsilon},
\end{equation}
where $\epsilon$ is a small constant to ensure numerical stability. This formulation captures the relative dominance of high-frequency components, which typically decrease in adversarially smoothed images \cite{frequency3}.

\subsubsection{Spectral Entropy Surrogate}  \label{spectral_entropy}
To further characterize the frequency distribution, we compute the spectral entropy of the normalized magnitude spectrum \cite{frequency1}.  
Let
\begin{equation}
    q(u,v) = \frac{M(u,v)}{\sum_{u,v} M(u,v)},
\end{equation}
which forms a discrete probability distribution over the spectrum. Classical spectral entropy is defined in the Shannon form \cite{frequency1} as
$H = - \sum_{u=0}^{H-1} \sum_{v=0}^{W-1}q(u,v)\, \log q(u,v).$, which yields a non–negative value that increases
with the uniformity of the spectrum.  
In this work, we employ a sign-free variant defined as
\begin{equation}
    H_{\text{sur}} = \sum_{u=0}^{H-1} \sum_{v=0}^{W-1}q(u,v)\, \log q(u,v).
\end{equation}
Since $0 < q(u,v) \le 1$ implies $\log q(u,v) \le 0$, we have $H_{\text{sur}} \le 0$.
Although $H_{\text{sur}}$ differs from the Shannon form by a negative factor,
it preserves the same monotonic behavior: spectra with higher dispersion yield
values closer to zero, whereas spectra with strong low-frequency concentration
produce values with larger negative magnitudes.

\begin{figure*}[h]
\centering
\includegraphics[width=0.9\textwidth]{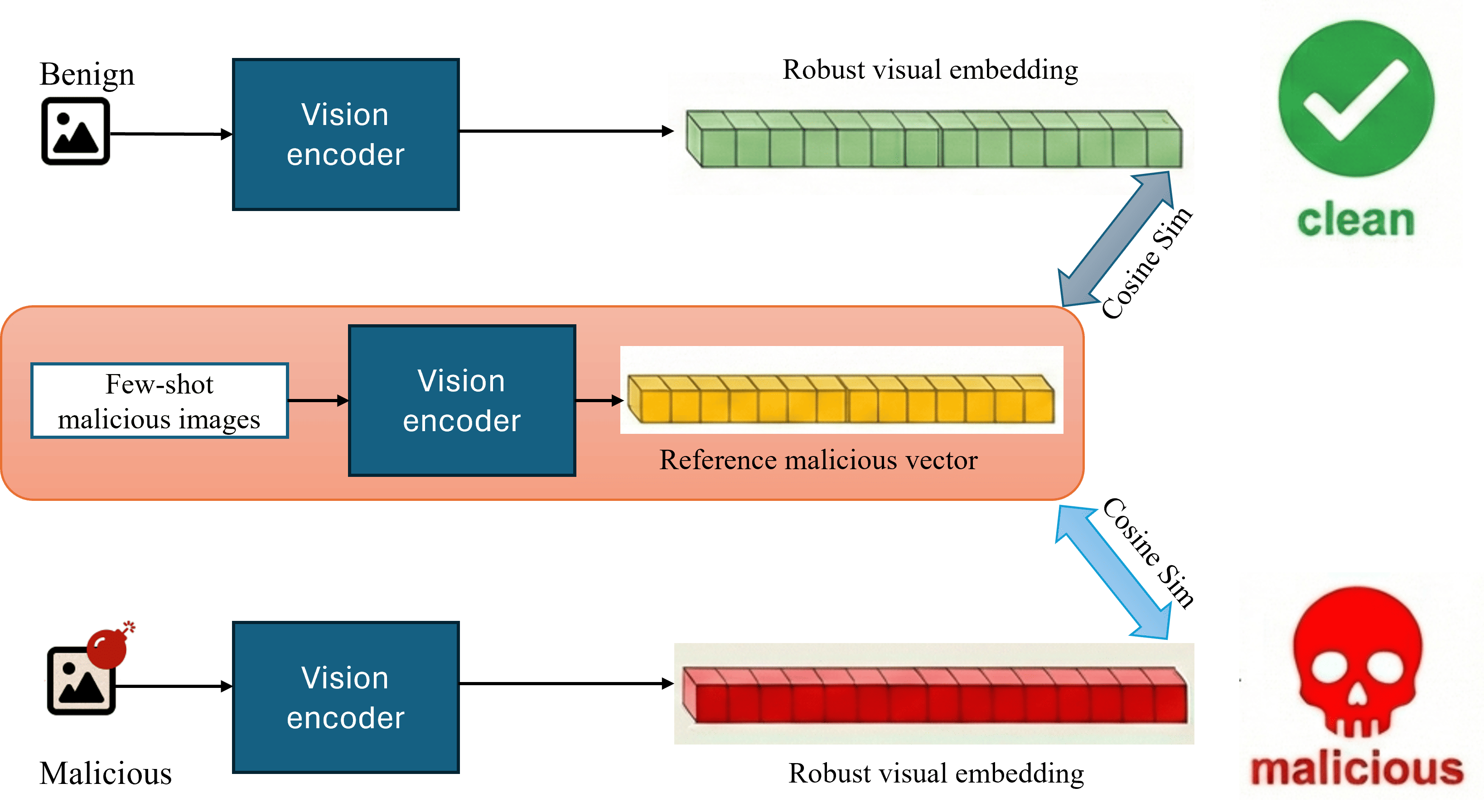}
\caption{Overview of the proposed white-box detector based on image vector embeddings.}
\label{whitebox_scheme}
\end{figure*}

\subsubsection{DFT-Based Malicious Score} \label{DFT_score}
To integrate the \emph{high-to-Low Frequency Energy Ratio} and the \emph{Spectral Entropy Surrogate} into a single indicator, we define the DFT-based malicious score as the average of the two metrics:
\begin{equation}
    S_{\text{DFT}} = \frac{1}{2}\left( H_{\text{HL}} + H_{\text{sur}} \right).
\end{equation}
This formulation provides a balanced measure reflecting both the concentration of high-to-Low frequency energy and the diversity of spectral components. A threshold applied to $S_{\text{DFT}}$ allows effective distinction between clean and malicious images.

\subsection{Proposed White-box Detector Based on image vector embeddings} \label{white_box}
As discussed in Section~\ref{VLM}, for a given visual input $I$, the visual encoder produces an embedding denoted by $h_e = F_e(I)$. This visual embedding strategy was previously introduced as a mechanism for steering the latent space to mitigate hallucination~\cite{steering}. We adopt this approach based on the observation that both hallucination and adversarial manipulation are closely related to instability in the model’s internal representations under input perturbations. In particular, hallucination arises when the model’s representations are overly sensitive to spurious or non-robust visual features, while malicious image prompt attacks deliberately exploit such sensitivities to inject adversarial signals.

To obtain a more robust representation of the visual input, we follow~\cite{steering} and compute the average embedding over a set of randomly masked variants of $I$.


\begin{equation}
    \bar{h}_e = \sum_{i=1}^{t} h_e^{\mathbf{C}_i(I)},
\end{equation}
where $\mathbf{C}_i(I)$ denotes the $i$-th random masking operation applied to the image, and $t$ indicates the total number of corrupted copies. The final robust visual embedding \cite{steering} is then defined as:
\begin{equation}
    h_r = \bar{h}_e - h_e,
\end{equation}
which characterizes the deviation induced by input perturbations and captures the stability of the visual encoder.

We compute the robust image vector embeddings under different few-shot settings $k \in [4, 32]$ for both clean and malicious images using the LLaVA-1.6-Vicuna-7B model~\cite{llava1}, as illustrated in Fig.~\ref{few_shot_output}. In this figure, the $x$-axis corresponds to the layer index of the visual encoder, whereas the $y$-axis denotes the mean activation values of the visual tokens. All results are averaged over five independent trials to mitigate randomness arising from masking and sampling procedures. The empirical trends exhibit a clear and stable separation between the embeddings of clean and malicious inputs, with the discrepancy further enlarging as $k$ increases. 

This behavior is expected, as increasing $k$ effectively averages over a larger set of randomly masked views, thereby reducing the variance of the estimated embedding and emphasizing features that are consistent across perturbations. For clean images, the underlying semantic content remains stable under masking, leading to highly consistent embeddings. In contrast, adversarial perturbations are typically non-robust and not semantically grounded, causing their influence to diminish under repeated masking and averaging. As a result, the aggregated embeddings of malicious inputs diverge more significantly from those of clean inputs as $k$ increases, leading to an amplified separation in the representation space.

This observation motivates the design of a white-box detection mechanism based on robust image vector embeddings. In particular, we introduce the \emph{Embedding-Based Malicious Score}, defined as follows.


\subsubsection{Embedding-Based Malicious Score}

To characterize the latent patterns of malicious image prompts, we collect a small few-shot set of adversarial examples 
\(
\{I_j^{\text{mal}}\}_{j=1}^{k}.
\)
Their corresponding robust image vector embeddings 
\(
\{h_r(I_j^{\text{mal}})\}_{j=1}^{k}
\)
are aggregated to form a reference malicious vector:
\begin{equation}
    {h_r}^{\text{ref}} = \frac{1}{k} \sum_{j=1}^{k} h_r(I_j^{\text{mal}}),
\end{equation}

Given a test image prompt $I_p$, we compute the cosine similarity between its robust visual embedding and the reference malicious vector:
\begin{equation}
    S_{\text{RVE}} = 
    \cos\!\left(h_r(I_p),\, {h_r}^{\text{ref}}\right)
    = 
    \frac{h_r(I_p)^{\top}{h_r}^{\text{ref}}}
         {\|h_r(I_p)\|_2 \, \|{h_r}^{\text{ref}}\|_2 }.
\end{equation}

A higher similarity score indicates that the test image’s robust visual representation is more aligned with the malicious embedding vector. Therefore, the detector categorizes an image prompt as malicious if its similarity exceeds a threshold.

The overall architecture of the proposed white-box detector is illustrated in Fig. \ref{whitebox_scheme}. This white-box design introduces no additional training overhead, requires only a few malicious examples to construct the reference malicious vector, and directly exploits the latent space of the visual encoder for robust malicious prompt detection.

\subsection{Hybrid Detection Strategy} \label{hybrid}
While the proposed black-box Fourier-domain detector and white-box embedding-based detector each exhibit the capability to detect malicious image prompts, their detection signals stem from fundamentally different perspectives: the former exploits input-level spectral characteristics, whereas the latter captures latent-space deviations revealed through the behavior of the visual encoder. To harness the complementary strengths of these two modalities, we introduce a hybrid detection mechanism that integrates their respective scores. Specifically, the two detectors are combined to produce a unified malicious detection score for DE-FIVE:
\begin{equation}
S_{\text{VG}} = \alpha S_{\text{DFT}} + (1-\alpha) S_{\text{RVE}},
\end{equation}

where $\alpha \in \{0,1\}$ denotes a non-negative weighting coefficient. The proposed weighting formulation allows for an adjustable balance between robustness and sensitivity: increasing $\alpha$ prioritizes generalization at the input level, while increasing $1-\alpha$ accentuates model-dependent effects in the latent representation space.

\begin{table*}[h]
\centering
\caption{\textbf{Detection performance against meta-instruction prompts under different meta-objectives, measured by AUROC.} ``Training-Free'' indicates whether an approach requires a training phase.
\textbf{Bold} numbers denote the best performance.}
\label{main_meta}
\resizebox{\textwidth}{!}{
\begin{tabular}{llc|cccccc}
\toprule
Model & Method & Training-Free & Language & Politics & Formality & Spam & Sentiment & Average \\
\midrule
\multirow{6}{*}{LLaVA-1.6}
& GradSafe \cite{Gradsafe}      & \cmark  & 72.80 & 63.97 & 66.94 & 60.70 & 61.45 & 65.17 \\
& JailGuard \cite{jailguard}      & \xmark  & 67.94 & 68.23 & 71.00 & 61.27 & 64.36 & 66.56 \\
& MirrorCheck \cite{Mirrorcheck}   & \xmark  & 77.98 & 70.12 & 74.65 & 63.29 & 72.92 & 71.79 \\
& Perplexity \cite{perplexity}    & \xmark  & 71.82 & 79.27 & 62.34 & 92.36 & 92.52 & 79.66 \\
& VLMGuard \cite{Vlmguard}      & \xmark  & 94.27 & 88.24 & 90.29 & \textbf{96.21} & 95.38 & 92.87 \\
& DE-FIVE (Ours) & \cmark & \textbf{96.59} & \textbf{95.33} & \textbf{93.33} & 95.34 & \textbf{96.15} & \textbf{95.35} \\
\midrule
\multirow{6}{*}{Phi-3}
& GradSafe \cite{Gradsafe}        & \cmark  & 73.46 & 57.39 & 70.45 & 53.75 & 63.07 & 63.62 \\
& JailGuard \cite{jailguard}      & \xmark  & 72.67 & 74.48 & 75.29 & 70.38 & 66.24 & 71.81 \\
& MirrorCheck \cite{Mirrorcheck}   & \xmark  & 80.27 & 71.09 & 73.57 & 70.04 & 72.37 & 73.74 \\
& Perplexity \cite{perplexity}    & \xmark  & 89.89 & 84.62 & 87.13 & 89.94 & 88.08 & 87.93 \\
& VLMGuard \cite{Vlmguard}      & \xmark  & 94.31 & 92.20 & \textbf{98.75} & 93.04 & 81.28 & 92.11 \\
& DE-FIVE (Ours) & \cmark & \textbf{97.10} & \textbf{94.66} & 95.58 & \textbf{96.97} & \textbf{90.35} & \textbf{94.87} \\
\bottomrule
\end{tabular}
}
\end{table*}

\section{Experiments}

This section presents an evaluation of the proposed method for the task of malicious image prompt detection. We first detail the experimental setup in Section~\ref{setting}, and then report the main results along with a comprehensive analysis in Sections~\ref{main_result}–\ref{analysis}.

\subsection{Experiment Setup}  \label{setting}

\subsubsection{Datasets and Models} \label{dataset}

We evaluate the proposed approach under two representative categories of malicious image prompt attacks: \textbf{meta-instruction prompts} for indirect prompt injection attacks~\cite{injection4} and \textbf{jailbreak prompts} for jailbreak attacks~\cite{adversarial4}. These two categories are selected as they capture fundamentally different attack mechanisms, hidden instruction injection versus safety alignment bypass, thereby enabling a comprehensive evaluation of detection robustness.

\textbf{Meta-instruction prompts}: We adopt the dataset introduced in~\cite{Vlmguard}, which comprises 25 benign images from ImageNet, each paired with 60 questions. This dataset is chosen because it provides a controlled and diverse benchmark for evaluating indirect prompt injection, where malicious intent is embedded within the visual modality rather than explicitly expressed in text. Following~\cite{injection4}, a total of 300 malicious meta-instruction images are generated using 40 training question–answer pairs. These meta-instruction pairs target five distinct meta-objectives: LANGUAGE, POLITICS, FORMALITY, SPAM, and SENTIMENT~\cite{injection4}, ensuring coverage of diverse behavioral manipulations. Consequently, the test dataset consists of 25 benign and 300 malicious images as visual prompts, each paired with 20 textual questions, following the evaluation protocol in~\cite{Vlmguard}.

\textbf{Jailbreak prompts}: Malicious images are generated using the XSTest dataset~\cite{xstest}, which consists of 250 safe prompts and 200 crafted unsafe prompts, together with 200 benign images from ImageNet, following the procedure described in~\cite{adversarial4}. This dataset is selected because it is widely used for evaluating jailbreak attacks and provides a standardized benchmark for assessing a model’s robustness against safety violations. Consequently, the test dataset comprises 200 benign and 200 malicious images as visual prompts, each paired with 50 safe and 100 unsafe textual prompts, enabling systematic evaluation across both benign and adversarial query settings.

\textbf{Models}: We further evaluate the proposed approach on two families of vision–language models: LLaVA-1.6 (LLaVA-1.6-Vicuna-7B)~\cite{llava1} and Phi-3 (Phi-3-Vision)~\cite{vlm1}, following the experimental protocol in~\cite{Vlmguard}. These models are selected as they represent state-of-the-art open-source VLM architectures with different design paradigms and training strategies, allowing us to assess the generalizability of the proposed method across heterogeneous model backbones. In particular, evaluating across multiple model families ensures that the proposed detection framework does not overfit to a specific architecture and remains effective in diverse real-world deployment scenarios.

\subsubsection{Baselines} 
We evaluate our method against a diverse set of baseline approaches, including detection methods designed specifically for jailbreak prompts, such as Perplexity \cite{perplexity}, GradSafe \cite{Gradsafe}, MirrorCheck \cite{Mirrorcheck}, and JailGuard \cite{jailguard}, as well as VLMGuard \cite{Vlmguard}, which is capable of handling both meta-instruction and jailbreak prompts.

For all baselines, we follow the original papers and use their publicly available implementations. Specifically, we directly adopt the released code and pretrained models. For methods that require threshold-based decisions (e.g., perplexity or gradient-based scores), we use the default thresholds provided in the original implementations. All methods are evaluated under the same experimental setting and input data, ensuring a consistent and fair comparison across approaches.

\subsubsection{Evaluation Details and Metrics}  \label{evaluate_details}
To ensure a fair comparison, we evaluate both the baseline methods and our proposed approach on the same test dataset, as described in Section~\ref{dataset}. For the baseline methods, we adopt the default experimental configurations reported in their original papers. For the proposed DE-FIVE method, we set $r_{\text{low}} = 8$ and $\epsilon = 1\times10^{-8}$ in Section~\ref{frequancy_ratio}, and adopt a mask ratio of 0.99 with 50 random masks, following \cite{steering}, to compute robust image vector embeddings in Section~\ref{white_box}. In addition, we employ a few-shot setting with $k = 32$ to construct the reference malicious vector, as described in Section~\ref{white_box}. This choice reflects a trade-off between detection performance and data efficiency, as increasing $k$ provides more representative malicious features but also incurs higher data requirements. We further set the non-negative weighting coefficient $\alpha$ to 0.2 for meta-instruction prompts and 0.1 for jailbreak prompts, as described in Section~\ref{hybrid}. These values are selected based on their empirical performance, where they achieve the best overall detection results. We also include a comprehensive ablation study in Section~\ref{analysis} to analyze the impact of both $k$ and $\alpha$, which further justifies these configurations for the main result comparisons.

Following \cite{perplexity,Gradsafe,Vlmguard}, we use the area under the receiver operating characteristic curve (AUROC) as the evaluation metric, which quantifies the performance of a binary classifier across varying thresholds. The final results are averaged over five runs.

\subsection{Main Result}  \label{main_result}

\textbf{Detection performance against meta-instruction prompts.} The experimental results reported in Table~\ref{main_meta} demonstrate that the proposed training-free DE-FIVE consistently achieves robust performance in detecting meta-instruction prompts across diverse meta-objectives. Moreover, the proposed approach achieves the highest average AUROC scores on both the LLaVA-1.6 and Phi-3 models when compared with all baseline methods.

Among the baseline approaches, GradSafe~\cite{Gradsafe} is the only training-free method; however, it achieves the lowest detection performance. The remaining four baseline methods~\cite{jailguard,Mirrorcheck,perplexity,Vlmguard} demonstrate improved detection performance relative to GradSafe, but all rely on a training stage. In contrast, compared with the state-of-the-art VLMGuard~\cite{Vlmguard}, which requires substantial amounts of unlabeled data for training, the proposed training-free DE-FIVE achieves average AUROC improvements of 2.6\% and 2.9\% on the LLaVA-1.6 and Phi-3 models, respectively.

\textbf{Detection performance against jailbreak prompts.} Beyond meta-instruction prompts, the experimental results reported in Table~\ref{main_jailbreak} indicate that the proposed training-free DE-FIVE also achieves the highest detection performance against jailbreak prompts on both the LLaVA-1.6 and Phi-3 models. The consistently strong performance across different VLM backbones suggests that DE-FIVE operates effectively without any training stage, making it a practical and easily deployable solution for real-world VLM safety applications.

\begin{table}[t]
\centering
\caption{\textbf{Detection performance against jailbreak prompts, measured by AUROC.} \textbf{Bold} numbers denote the best performance.}
\label{main_jailbreak}
\begin{tabular}{ll|c}
\toprule
Model & Method & Jailbreak Prompts \\
\midrule
\multirow{6}{*}{LLaVA-1.6}
& Perplexity \cite{perplexity}         & 69.31 \\
& MirrorCheck \cite{Mirrorcheck}         & 83.26 \\
& JailGuard \cite{jailguard}           & 88.49 \\
& GradSafe \cite{Gradsafe}            & 91.09 \\
& VLMGuard \cite{Vlmguard}            & 94.27 \\
& DE-FIVE (Ours)  & \textbf{95.22} \\
\midrule
\multirow{6}{*}{Phi-3}
& MirrorCheck \cite{Mirrorcheck}        & 73.86 \\
& GradSafe \cite{Gradsafe}            & 82.77 \\
& JailGuard \cite{jailguard}           & 82.91 \\
& Perplexity \cite{perplexity}         & 94.36 \\
& VLMGuard \cite{Vlmguard}            & 95.74 \\
& DE-FIVE (Ours)  & \textbf{96.03} \\
\bottomrule
\end{tabular}
\end{table}

\begin{table*}[h]
\centering
\caption{Ablation study on the proposed black-box detector and jailbreak defense baselines under meta-instruction and jailbreak prompts, measured by AUROC. \textbf{Bold} numbers denote the best performance.}
\label{fig_black_box}
\resizebox{\textwidth}{!}{
\begin{tabular}{l|cccccc|c}
\toprule
\multirow{2}{*}{Method} 
& \multicolumn{6}{c|}{Meta-instruction prompts} 
& \multirow{2}{*}{Jailbreak prompts} \\
\cmidrule(lr){2-7}
& Language & Politics & Formality & Spam & Sentiment & Average &  \\
\midrule
GradSafe  \cite{Gradsafe}       & 72.80 & 63.97 & 66.94 & 60.70 & 61.45 & 65.17 & \textbf{91.09} \\
JailGuard \cite{jailguard}      & 67.94 & 68.23 & 71.00 & 61.27 & 64.36 & 66.56 & 88.49 \\
MirrorCheck \cite{Mirrorcheck}   & 77.98 & 70.12 & 74.65 & 63.29 & 72.92 & 71.79 & 83.26 \\
Perplexity \cite{perplexity}    & 71.82 & 79.27 & 62.34 & \textbf{92.36} & \textbf{92.52} & 79.66 & 69.31 \\
\midrule
(1) High-to-Low Frequency Energy Ratio             & 67.26 & 66.44 & 66.45 & 67.55 & 65.92 & 66.72 & 65.01 \\
(2) Spectral Entropy Surrogate             & 84.18 & 84.00 & 82.22 & 82.23 & 83.28 & 83.18 & 84.26 \\
(3) DFT-based malicious scores           & \textbf{86.72} & \textbf{86.66} & \textbf{86.22} & 86.67 & 86.33 & \textbf{86.52} & 87.25 \\
\bottomrule
\end{tabular}
}
\end{table*}

\begin{table*}[h]
\centering
\caption{Ablation study on the proposed white-box detector and VLMGuard baseline under meta-instruction and jailbreak prompts, measured by AUROC. \textbf{Bold} numbers denote the best performance.}
\label{fig_white_box}
\resizebox{\textwidth}{!}{
\begin{tabular}{l|cccccc|c}
\toprule
\multirow{2}{*}{Method}
& \multicolumn{6}{c|}{Meta-instruction prompts}
& \multirow{2}{*}{Jailbreak prompts} \\
\cmidrule(lr){2-7}
& Language & Politics & Formality & Spam & Sentiment & Average & \\
\midrule
VLMGuard \cite{Vlmguard}
& \textbf{94.27} & 88.24 & 90.29 & 96.21 & \textbf{95.38} & 92.87 & 94.27 \\
(1) proposed white-box ($k=32$)
& 90.52 & 95.56 & 89.78 & 92.89 & 91.85 & 92.19 & 92.89 \\
(2) proposed white-box ($k=64$)
& 90.82 & \textbf{98.89} & \textbf{94.67} & \textbf{96.89} & 92.30 & \textbf{94.71} & \textbf{95.02} \\
\bottomrule
\end{tabular}
}
\end{table*}

\begin{table*}[h]
\centering
\caption{Ablation study on weighting coefficients under meta-instruction and jailbreak prompts, measured by AUROC. \textbf{Bold} numbers denote the best performance.}
\label{weighting_coeff}
\resizebox{\textwidth}{!}{
\begin{tabular}{c|cccccc|c}
\toprule
\multirow{2}{*}{Weighting coefficients}
& \multicolumn{6}{c|}{Meta-instruction prompts}
& \multirow{2}{*}{Jailbreak prompts} \\
\cmidrule(lr){2-7}
& Language & Politics & Formality & Spam & Sentiment & Average & \\
\midrule
0.1 & 95.71 & \textbf{96.89} & 92.00 & 95.33 & 94.96 & 94.98 & \textbf{95.22} \\
0.2 & \textbf{96.59} & 95.33 & \textbf{93.33} & \textbf{95.34} & \textbf{96.15} & \textbf{95.35} & 93.01 \\
0.3 & 95.40 & 93.33 & 92.44 & 94.22 & 94.22 & 93.92 & 91.05 \\
0.4 & 94.07 & 92.67 & 91.78 & 92.89 & 93.04 & 92.89 & 90.79 \\
0.5 & 92.74 & 91.78 & 90.22 & 92.00 & 92.44 & 91.84 & 90.49 \\
0.6 & 90.81 & 91.11 & 89.56 & 91.33 & 92.00 & 90.96 & 89.52 \\
0.7 & 89.48 & 89.33 & 88.89 & 89.11 & 89.63 & 89.29 & 88.60 \\
0.8 & 88.00 & 88.67 & 87.33 & 88.22 & 88.30 & 88.10 & 88.05 \\
0.9 & 87.26 & 87.56 & 86.44 & 86.89 & 87.55 & 87.14 & 87.55 \\
\bottomrule
\end{tabular}
}
\end{table*}

\subsection{Analysis}  \label{analysis}
In this section, we present additional analyses and ablation studies to evaluate the effectiveness of the proposed DE-FIVE. All experiments are conducted on the LLaVA-1.6 model.

\textbf{Ablation on the proposed black-box detector:}
To investigate the effectiveness of the proposed black-box detector when operating on different Fourier-domain features, we compare three configurations:
(1) the High-to-Low Frequency Energy Ratio described in Section~\ref{frequancy_ratio},
(2) the Spectral Entropy Surrogate described in Section~\ref{spectral_entropy}, and
(3) the proposed DFT-based malicious scores described in Section~\ref{DFT_score}.
We further compare these configurations with existing jailbreak defense baselines, including Perplexity~\cite{perplexity}, GradSafe~\cite{Gradsafe}, MirrorCheck~\cite{Mirrorcheck}, and JailGuard~\cite{jailguard}.

As shown in Table~\ref{fig_black_box}, Perplexity \cite{perplexity} achieves the highest AUROC against meta-instruction prompts but exhibits the lowest AUROC against jailbreak prompts. This observation indicates that existing jailbreak defense baselines struggle to maintain robust performance across both meta-instruction and jailbreak prompt settings. In contrast, the proposed DFT-based malicious scores achieve consistently strong performance on both types of prompts.

Moreover, the DFT-based malicious scores outperform detectors based on a single Fourier-domain feature, such as (1) the High-to-Low Frequency Energy Ratio or (2) Spectral Entropy Surrogate alone. Notably, the proposed approach achieves performance comparable to state-of-the-art jailbreak defense baselines while incurring no additional computational or training cost.

\textbf{Ablation on the proposed white-box detector:}
To demonstrate the effectiveness of the proposed white-box detector under different few-shot settings $k$, we evaluate two configurations:
(1) the proposed white-box detector with $k = 32$, as described in Section~\ref{evaluate_details}, and
(2) the proposed white-box detector with $k = 64$. These two configurations are chosen to reflect a meaningful trade-off between data efficiency and detection performance. Specifically, $k = 32$ represents a data-efficient setting with limited access to malicious examples, while $k = 64$ corresponds to a higher-resource scenario where more samples are available to construct a more representative malicious reference vector. Comparing these two settings allows us to assess how the proposed method scales with additional data and whether performance gains justify the increased data requirement.

We further compare these configurations with the best-performing baseline, VLMGuard~\cite{Vlmguard}.

As reported in Table~\ref{fig_white_box}, VLMGuard~\cite{Vlmguard} achieves strong performance across both meta-instruction and jailbreak prompt scenarios. The proposed white-box detector with $k = 32$ attains slightly lower AUROC than VLMGuard. Notably, increasing the few-shot setting to $k = 64$ enables the proposed detector to surpass the VLMGuard baseline in terms of AUROC performance.

However, we adopt $k = 32$ as the default setting in the main experiments, as it achieves competitive performance while requiring significantly fewer malicious samples. This choice reflects a more practical and data-efficient scenario, where access to large numbers of high-quality malicious examples may be limited.

\textbf{Ablation on the non-negative weighting coefficient $\alpha$:}
In Table~\ref{weighting_coeff}, we ablate the effect of the non-negative weighting coefficient $\alpha$ under both meta-instruction and jailbreak prompt settings. As described in Section~\ref{hybrid}, the choice of $\alpha$ directly influences the hybrid detection performance by assigning different weights to the embedding-based malicious score (i.e., the proposed white-box detector) and the DFT-based malicious score (i.e., the proposed black-box detector).

The results indicate that meta-instruction prompts achieve the best average performance when $\alpha = 0.2$, whereas jailbreak prompts obtain optimal performance when $\alpha = 0.1$. A lower value of $\alpha$ assigns greater weight to the embedding-based malicious score and correspondingly less weight to the DFT-based malicious score. Overall, these findings suggest that allocating more weight to the proposed white-box detector generally leads to improved performance of the final DE-FIVE.

\section{Conclusions}
Existing approaches for detecting malicious prompts primarily focus on text-based jailbreak attacks or rely on training-heavy pipelines, making them less effective or inefficient when extended to multimodal settings. In particular, they often struggle to generalize across different types of malicious image prompts, such as meta-instruction and jailbreak prompts, or require substantial labeled data and retraining to adapt to new attack patterns.

To address these limitations, we introduce a novel training-free framework, termed \textsc{DE-FIVE}, for detecting malicious image prompts, including both meta-instruction and jailbreak prompts in this paper. Specifically, we first propose a black-box detector based on two complementary Fourier-domain features. We then develop a white-box detector leveraging image vector embeddings. By assigning adaptive weights to these two detectors, we further construct a hybrid detection strategy that integrates their respective strengths.

Experimental results demonstrate that \textsc{DE-FIVE} achieves superior detection performance without incurring any additional training cost when identifying malicious image prompts. Furthermore, comprehensive analyses and ablation studies provide deeper insights into the effectiveness and robustness of the proposed \textsc{DE-FIVE} framework.

%
%
%



\end{document}